\documentclass[aps,prc,preprint,amssymb,amsmath,superscriptaddress,floatfix,10pt]{revtex4-1}
\usepackage{mathptmx}
\usepackage{graphicx}
\usepackage{color}  
\makeindex
\newcommand{\textblue}{\textcolor[rgb]{0.00,0.07,1.00}}

\begin{document}

\title{Some properties of plasma surrounding brown dwarfs}
\date{2023-10-26}
\author{Dmitry Kobyakov}
\email{dmitry.kobyakov@appl.sci-nnov.ru}
\affiliation{Institute of Applied Physics of the Russian Academy of Sciences, 603950 Nizhny Novgorod, Russia}

\begin{abstract}
Recently, brown dwarfs have emerged as a new topic for the astrophysical studies. These objects are intermediate between solar-type stars and giant gaseous planets. In this article, the analogies between brown dwarfs and the planet Jupiter are considered with a focus on the surrounding plasma. I consider the magnetohydrodynamic version of the Rayleigh-Taylor instability (or so called ``interchange instability'') as a minimal model of the expansion of the plasma disc surrounding Jupiter. By comparing the theoretical prediction for the radial expansion rate of the disc with the observations I quantitatively confirm the existing qualitative result, which predicts that the Rayleigh-Taylor instability provides too quick expansion. Therefore, in the realistic plasma disc yet another mechanism must operate which slows down the expansion. I suggest that similar mechanisms take place in the observed radiation belts of brown dwarfs.
\end{abstract}

\maketitle

\emph{Introduction.}
Brown dwarf is a stellar-type celestial body with mass $M_{*}$ in the range $13M_{\mathrm{Jup}}<M_{*}<80M_{\mathrm{Jup}}$, or, in solar masses, $1.241\times10^{-2}M_{\bigodot}<M_{*}<7.636\times10^{-2}M_{\bigodot}$, where the lower limit corresponds to the minimum mass suitable for the stellar deiterium combustion and the upper limit corresponds to the minimum mass suitable for the stellar hydrogen combustion.
Here, $M_{\mathrm{Jup}}=1.8913\times10^{30}$ g is the Jupiter mass.
The spectral type of brown dwarf is in the range M7-M9, L, T, Y.
Its temperature is between 300 and 2500 K.
The dipolar magnetic field on the surface is typically of the order of $10^3-10^4$ G.
The possible emission types are radio, infrared, optical, ultravioler and X-ray \cite{Burrows2001,Hallinan2006,ZaitsevStepanov2022,Climent2023,Kao2023,Bespalov2018}.

Observations \cite{Climent2023,Kao2023} of the brown dwarf 2MASS J18353790+3259545 (equivalently denoted as LSR J1835+3259) with mass $\sim77M_{\mathrm{Jup}}$, radius $\sim1.07R_{\mathrm{Jup}}$ and rotation period $1.008\times10^{4}$ s, have revealed a radiation belt surrounding the star.
The radiation belt has radius $\sim17R_{\mathrm{Jup}}$, where $R_{\mathrm{Jup}}=7.1492\times10^{9}$ cm is Jupiter's radius \cite{Climent2023}.
The existence of the radiation belt, relatively strong magnetic field and rapid rotation observed from LSR J1835+3259 indicates that there are analogies between the radio emission mechanisms in its magnetosphere and the physics of the radiation belt of Jupiter.
At present, the origin of the plasma in the radiation belt of LSR J1835+3259 is unclear but it is likely that in analogy with the Jupiter-Io system there is a planetary satellite \cite{Climent2023}.

An elementary physical picture of the radiation belt is based on the model of the uniform (solid-like) rotation of the magnetosphere.
The mechanism maintaining the rotation of the plasma surrounding a rotating magnetic dipole with electrically conducting surface has been considered in \cite{HonesBergeson1965}.
The Alfven radius defines the radial distance from the star center to the point where the configuration of the magnetic field lines changes from closed to open (Fig. 1).
The black dot in Fig. 1 is the source of plasma (Io in case of Jupiter's magnetosphere).
\begin{figure}
\includegraphics[width=3.5in]{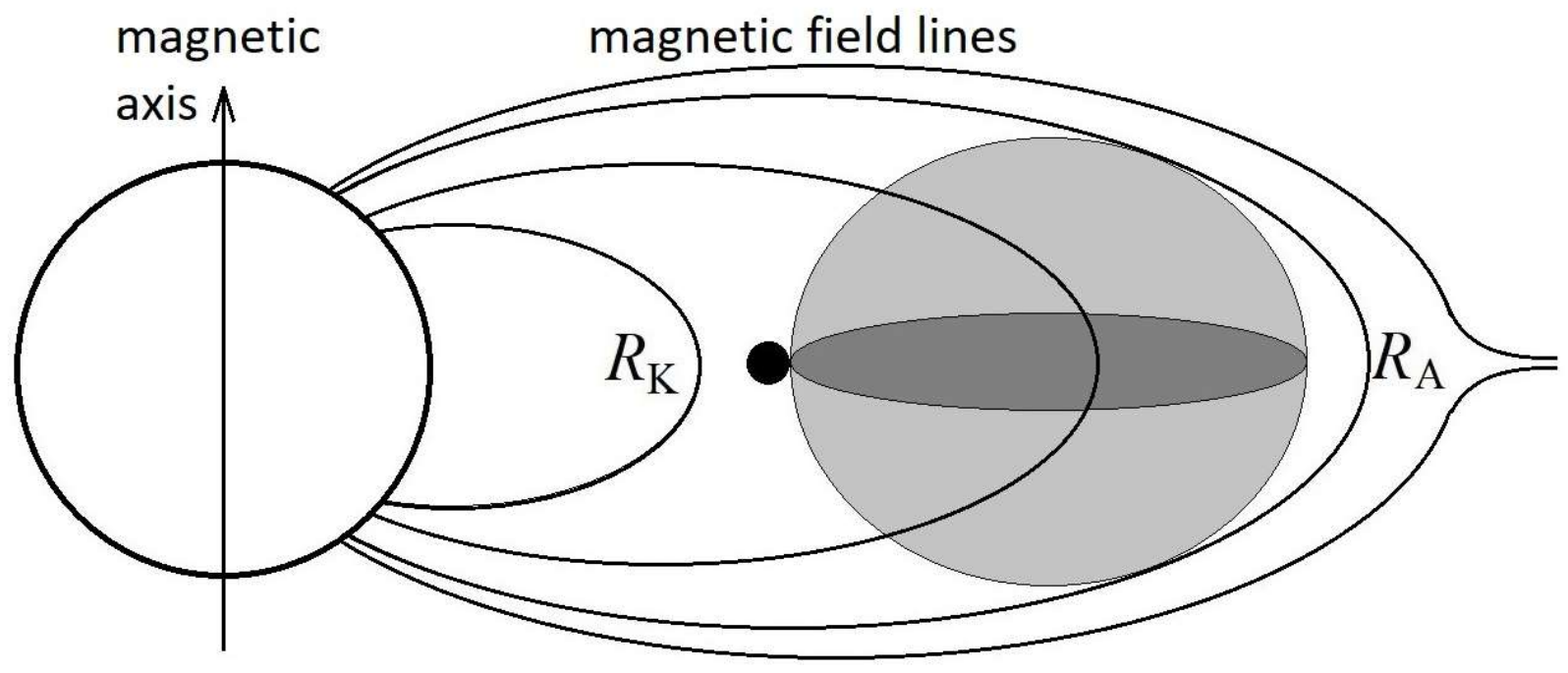}
\caption{}
\end{figure}

With $R_{\mathrm{K}}<R_{\mathrm{A}}$, the magnitosphere is centrifugal \cite{udDoulaOwocki2002}, where $R_{\mathrm{K}}=(GM_{*}/\Omega^2)^{1/3}$ is the Kepler radius (Fig. 1), $\Omega$ is the rotational angular frequency.
Formation of a plasma disc (Figs. 1,2) as a result of the magnetosphere rotation has been first shown for the magnetic star $\sigma$ Ori E \cite{Nakajima1985}.
The same mechanism leads to the formation of Jupiter's plasma disc.
The standard model of the radial expansion of Jupiter's plasma disc is the convective (or so called interchange) plasma instability of the plasma disc \cite{BagenalDols2020}.
However, there remains an open question \cite{BagenalDols2020}: \emph{why is the observed expansion of the plasma disc is significantly slower than the expansion rate predicted theoretically in the framework of the interchange plasma instability?}

\begin{figure}
\includegraphics[width=2.5in]{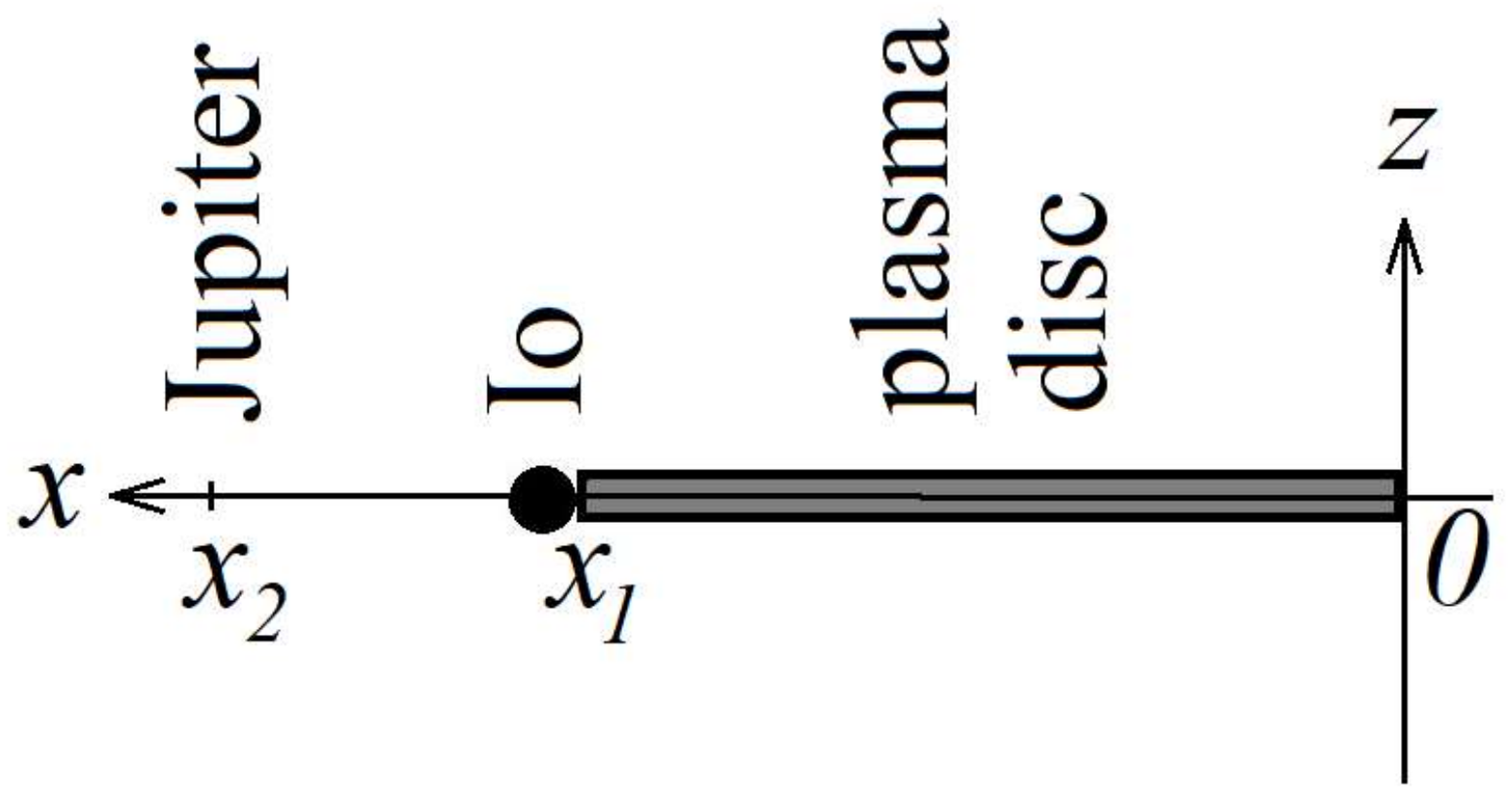}
\caption{}
\end{figure}
The notion of the ``interchange mode'' has appeared in the beginning of studies of the laboratory plasma.
It implies that the plasma and the confining magnetic field switch their spatial locations as a result of action of the external forces.
In dealing with the interchange plasma instability I will follow the book \cite{Goedbloed2019}.
The problem of the interchange plasma instability is analogous to the Rayleigh-Taylor instability.

Figure 3 shows a schematic picture of the plasma slab in a uniform external force field supported by the magnetic field.
Linearization of the equations of motion of ideal isothermic plasma with a perturbation of a fluid element $\pmb{\xi}$, the external free-fall acceleration $\mathbf{g}=(-g,0,0)$ and the perturbation wave vector $\mathbf{k_0}=(0,k_y,k_z)$, (Fig. 3), leads to the resulting potential energy $W$ of the system:
\begin{equation}\label{W}
  W=\frac{\xi_x(0)}{2k_0}\left[\frac{\left(\mathbf{k_0}\cdot\mathbf{B_0}\right)^2}{\tanh{k_0a}} - \rho_0k_0g + \frac{\left(\mathbf{k_0}\cdot\mathbf{\hat{B}_0}\right)^2}{\tanh{k_0b}}\right].
\end{equation}
Equation (\ref{W}) shows that (i) the external force $\mathbf{g}$ ($g\geq0$) always destabilizes the plasma, (ii) the magnetic induction may stabilize the plasma.

In case when the plasma is inhomogeneous along $x$ axis, the instability is described by the equation found for the first time in \cite{Goedbloed1971}.
If the conditions $\mathbf{B_0}\times\mathbf{\hat{B}_0}=0$ and $\mathbf{B_0}\cdot\mathbf{\hat{B}_0}>0$ are satisfied, the dispersion equation has the form
\begin{equation}\label{Dispersion}
  \omega^4-\Omega_1^4+\Omega_2^4=0,
\end{equation}
where $\Omega_1^4=\frac{b^2+2c^2}{b^2+c^2}k_{\parallel}^2b^2+\frac{k_0^2}{k_0^2+q^2}N_m^2$; $\Omega_2^4=\frac{c^2}{b^2+c^2}k_{\parallel}^2b^2\left(k_{\parallel}^2b^2 + \frac{k_0^2}{k_0^2+q^2}N_B^2\right)$; $c=\gamma[p(x=0)]/[\rho(x=0)]$; $b=B_0/\sqrt{\rho(x=0)}$; $\gamma$ is the adiabatic index; $k_{\parallel}$ is the component of $\mathbf{k_0}$ which is parallel to $\mathbf{B_0}$; $\xi\sim e^{iqx}$, $qL\gg1$, $L=(p+B^2/2)/\rho g$ is the size of equilibrium variations.
The frequencies (Brunt-V$\mathrm{\ddot{a}}$is$\mathrm{\ddot{a}}$l$\mathrm{\ddot{a}}$a and its magnetic modification \cite{Goedbloed2019}) are given by \begin{equation}\label{Frequencies}
  N_b^2=-\frac{1}{\rho}\left(\rho' g + \frac{\rho^2g^2}{\gamma p}\right),\quad N_m^2=-\frac{1}{\rho}\left(\rho' g + \frac{\rho^2g^2}{\gamma p+B^2}\right),
\end{equation}
where $\rho'\equiv\partial_x\rho|_{x=0}$.
The relation between the growth rates is defined by four quantities:
\begin{equation}\label{Gammas}
  \Gamma=-\frac{\rho'}{\rho}g,\quad \Gamma_B=\frac{\rho g^2}{\gamma p}, \quad \Gamma_m=\frac{\rho g^2}{\gamma p+B^2},\quad\Gamma_0=\frac{\Gamma_m^2}{\Gamma_B}.
\end{equation}
It has been known that (i) the plasma is stable when $\Gamma_B\leq\Gamma$; (ii) at $\Gamma_0\leq\Gamma<\Gamma_B$ the most unstable mode is the quasiinterchange mode ($k_{\parallel}\neq0$) and its growth rate is $\omega^2=-\frac{\rho g^2}{B^2}(1-\sqrt{\Gamma/\Gamma_B})^2$; at $\Gamma\leq\Gamma_0$ the most unstable is the interchange mode with the growth rate $\omega^2=\Gamma-\Gamma_m$.

\begin{figure}
\includegraphics[width=3.5in]{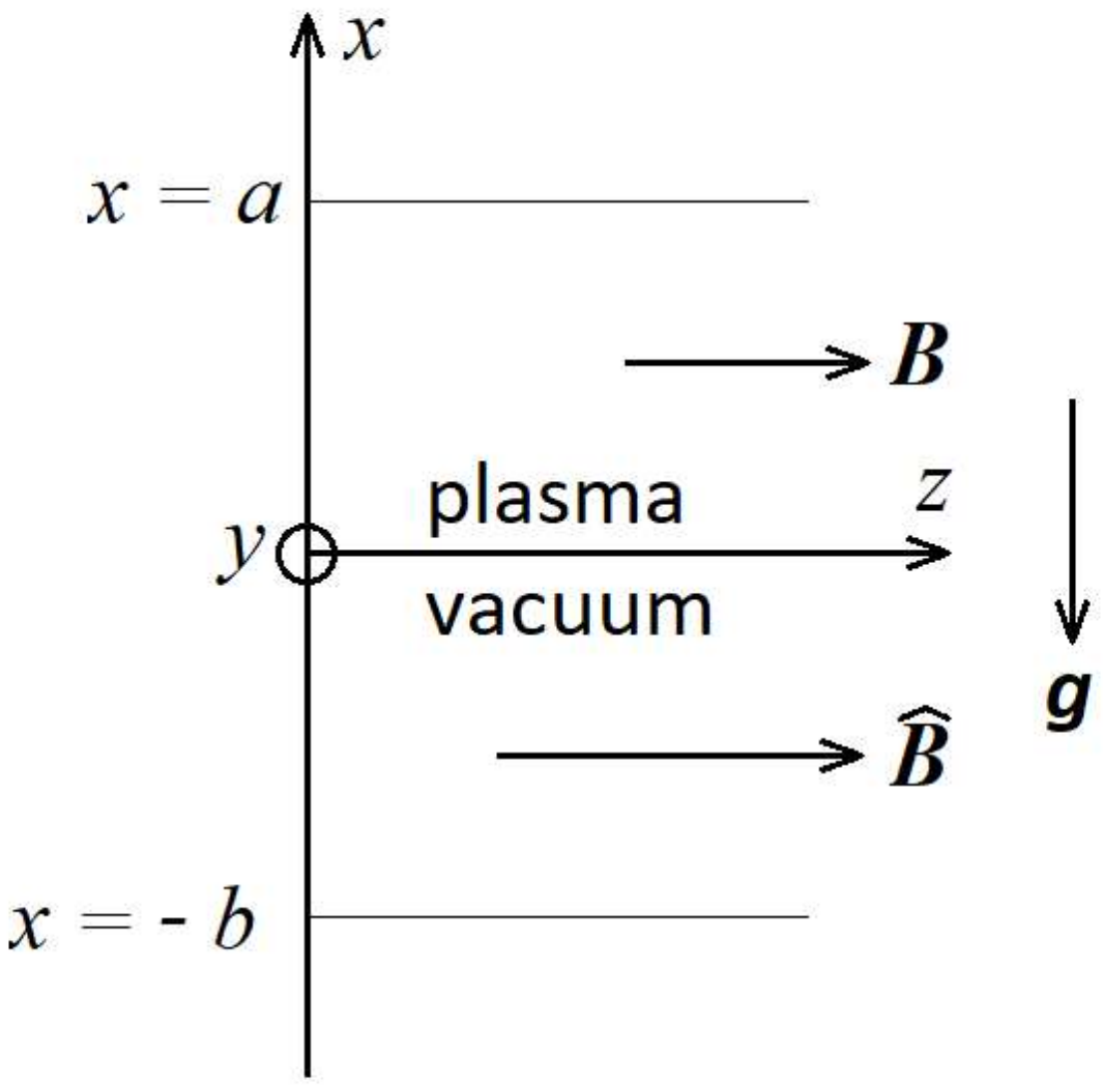}
\caption{}
\end{figure}
\emph{Numerical results.}
For Jupiter's plasma disc, the parameters entering Eq. (\ref{Dispersion}) are known from observations, and thus, the most unstable mode can be easily found.
Using figure 4 of \cite{Bespalov2006} I find the characteristic distance of the outer edge of the plasma disc ($x=0$) from Jupiter's center ($x=x_2$) (Fig. 2):
\begin{equation}\label{r2}
  x_2\approx20R_{\mathrm{Jup}}.
\end{equation}
The mass density of the electron-ion plasma $\rho=Am_pn(x=0)$, where $A\sim48$ is the atomic mass (assuming that the sulfur oxide is the ion component of plasma), $m_p$ is the proton mass, $n\propto(x_2-x)^{-3}$, from figure 4 of \cite{Bespalov2006}
\begin{eqnarray}\label{nr2}
  && n(x=0)\approx1\;\mathrm{cm}^{-3},\\
  && g\approx r_2\Omega_{\mathrm{Jup}}^2=4.422\times10^{3}\;\mathrm{cm}\,\mathrm{s}^{-2}
\end{eqnarray}
where $\Omega_{\mathrm{Jup}}=1.759\times10^{-4}$ rad $\mathrm{s}^{-1}$.
From these parameters I find
\begin{eqnarray}
  \label{G1}&& \Gamma=-9.277\times10^{-8}\;\mathrm{s}^{-2},\quad\Gamma_B=6.821\times10^{-4}\;\mathrm{s}^{-2},\\
  \label{G2}&& \Gamma_m=9.9\times10^{-7}\;\mathrm{s}^{-2},\quad\Gamma_0=1.437\times10^{-9}\;\mathrm{s}^{-2}.
\end{eqnarray}
It follows from Eqs. (\ref{G1})-(\ref{G2}) that the case $\Gamma<\Gamma_0$ (since $\Gamma<0$) is realized.
Therefore, the expansion of the plasma disc of Jupiter should occur due to the interchange mode with the characteristic growth rate from Eq. (\ref{Dispersion}):
\begin{equation}\label{tau_theory}
  \tau_{theory}\sim1.056\times10^{3}\;\mathrm{s}.
\end{equation}
This result implies that the theoretical prediction for the growth rate is significantly smaller than it is expected from observations.
The latter has the order of 20-80 days \cite{BagenalDols2020}, or in case of 20 days,
\begin{equation}\label{tau_obs}
  \tau_{observ}\sim1.728\times10^{6}\;\mathrm{s}.
\end{equation}

\emph{Conclusions.}
The quantitative estimate for the expansion rate of Jupiter's plasma disc, Eq. (\ref{tau_theory}), agrees with the qualitative prediction known from the literature \cite{BagenalDols2020}.
Specifically, the theory predicts a growth rate, Eq. (\ref{tau_theory}), which is a few orders of magnitude smaller than it is inferred from the observations, Eq. (\ref{tau_obs}).
In case when a brown dwarf possess a plasma disc, the analogous situation is expected.
Such a discrepancy between the theory and observations indicates that a significant piece of theoretical understanding of the plasma surrounding those celestial bodies is missing.
In the future work it is therefore necessary to identify possible physical mechanisms, which are responsible for the practical increase of the duration of the loss of matter.
It is necessary to analyze the following possible reasons.
(i) Nonzero shear of the magnetic field, which has not been included in the linear analysis in Eq. (\ref{Dispersion}).
(ii) Account for the Birkeland currents and the corresponding electric current in the plasma disc.
(iii) The action of the Kelvin-Helmholtz instability on the nonlinear stage of the interchange instability found in Eq. (\ref{Dispersion}).

\emph{Acknowledgements.} I thank P. A. Bespalov for helpful comments and discussions. This research was supported by the Russian Science Foundation under grant No. 20-12-00268.

\emph{Translated by the author.}

\end{document}